 \documentclass[prb,twocolumn,showpacs]{revtex4}
\usepackage{graphicx}
\usepackage{subfigure}
\usepackage{amssymb,amsmath}
\begin{document}
\bibliographystyle{apsrev}

\title{First principles investigations of the electronic, magnetic and 
       chemical bonding properties of $ {\bf CeTSn} $ 
       ($ {\bf T=Rh,Ru} $)}

\author{S.\ F.\ Matar,$^a$
        J.\ F.\ Riecken,$^b$
        B.\ Chevalier,$^a$
        R.\ P\"ottgen,$^b$ and  
  V.\ Eyert$^c$\footnote{Corresponding author: eyert@physik.uni-augsburg.de}}
\affiliation{$^a$ICMCB, CNRS, Universit\'e Bordeaux 1, 
             87 avenue du Docteur Albert Schweitzer, 
             33608 Pessac Cedex, France, \\
             $^b$Institut f\"ur Anorganische und Analytische Chemie, 
             Universit\"at M\"unster, 
             Corrensstrasse 30, 48149 M\"unster, Germany, \\
             $^c$Chemische Phsik und Materialwissenschaften, 
             Center for Electronic Correlations and Magnetism, 
             Institut f\"ur Physik, Universit\"at Augsburg, 
             86135 Augsburg, Germany}

\date{\today}

\pacs{07.55.Jg, 71.20.-b, 71.23}


\begin{abstract}
The electronic structures of $ {\rm CeRhSn} $ and $ {\rm CeRuSn} $ are 
self-consistently calculated within density functional theory using the 
local spin density approximation for exchange and correlation. In 
agreement with experimental findings, the analyses of the electronic 
structures and of the chemical bonding properties point to the absence 
of magnetization within the mixed valent Rh based system while a finite 
magnetic moment is observed for trivalent cerium within the Ru-based 
stannide, which contains both trivalent and intermediate valent Ce.
\end{abstract}
\maketitle

\section{Introduction}
Equiatomic cerium-transition metal (T)-stannides $ {\rm CeTSn} $ have 
intensively been studied over the past twenty years with respect to 
their outstanding magnetic properties. $ {\rm CeNiSn} $, \cite{takabatake87} 
$ {\rm CeRhSn} $, \cite{schmidt05} and $ {\rm CeIrSn} $ \cite{chevalier06} 
are intermediate valence systems, while $ {\rm CePdSn} $ 
($ T_N = 7.5 $\,K) \cite{adroja88} and $ {\rm CePtSn} $ ($ T_N = 8 $\,K) 
(see Ref.\ \onlinecite{riecken05} and references therein) are antiferromagnetic 
Kondo lattices. $ {\rm CeAgSn} $ orders antiferromagnetically at 6.5\,K
\cite{baran97} and $ {\rm CeAuSn} $ ($ T_C = 4.1 $\,K) \cite{lenkewitz96} 
is the only ferromagnet within this series. Especially $ {\rm CeNiSn} $, 
$ {\rm CePdSn} $, and $ {\rm CePtSn} $ have been thoroughly investigated.  
\cite{scifinder}  All these ternary stannides exhibit only one 
crystallographic cerium site in their structures. In that view, 
$ {\rm CeNiSn} $, $ {\rm CeRhSn} $, and $ {\rm CeIrSn} $ can be considered 
as homogeneous intermediate-valent systems. This is different in the 
recently reported stannide $ {\rm CeRuSn} $, \cite{riecken07} which is 
the first $ {\rm CeTSn} $ compound, which adopts a superstructure 
(at room temperature) with two crystallographically independent cerium 
sites, one trivalent and one intermediate-valent one. Below room 
temperature the structure becomes modulated and these modulations have 
clear consequences on the temperature dependence of the magnetic 
susceptibility and the specific heat. \cite{hoffmann07} Further, the 
system is found to order antiferromagnetically at $ T_N = 3.0(2) $\,K. 
Due to the peculiar structural behavior and the course of the physical 
properties we were interested in the electronic structure and chemical 
bonding properties of $ {\rm CeRuSn} $ in comparison to 
intermediate-valent $ {\rm CeRhSn} $. Our investigation is carried out 
in the framework of density functional theoretical (DFT)
\cite{hohenberg64,kohn65} and uses the scalar-relativistic 
implementation of the augmented spherical wave (ASW) method. 
\cite{wkg,revasw,bookasw}

\section{Crystal structures}
\begin{figure}[htbp]
\includegraphics[width=\columnwidth]{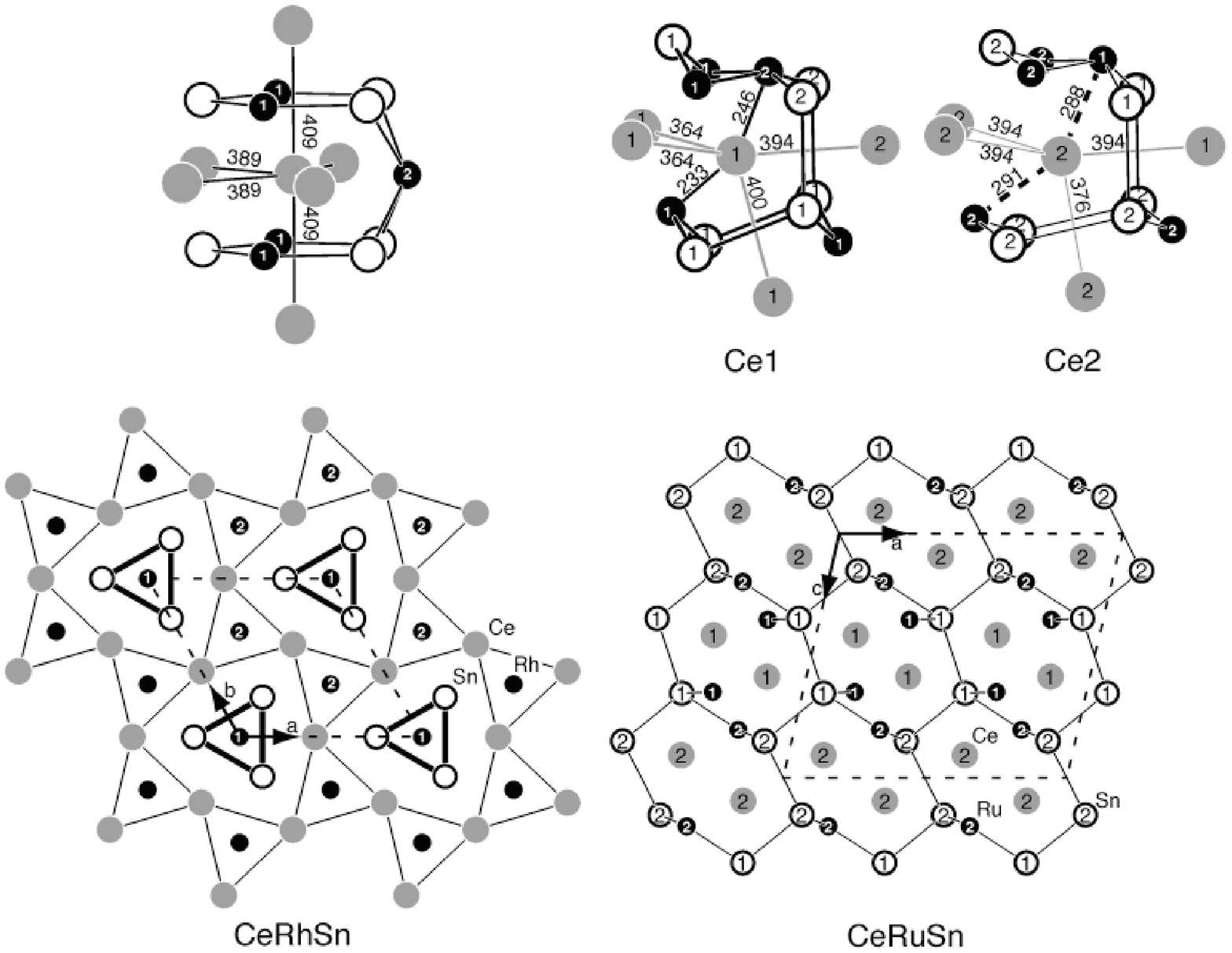}
\caption{The crystal structures of $ {\rm CeRhSn} $ (space group 
         $ P\bar{6}2m $) and $ {\rm CeRuSn} $ (space group $ C2/m $) 
         as projections along the short unit cell axes (bottom). 
         The cerium, rhodium (ruthenium), and tin atoms are drawn as 
         medium grey, black, and open circles, respectively. The 
         trigonal prismatic units in $ {\rm CeRhSn} $ and the 
         three-dimensional $[RuSn]$ network of $ {\rm CeRuSn} $ are 
         emphasized. The coordination polyhedra of the different 
         cerium sites are presented at the top of the drawing together 
         with relevant interatomic distances in units of pm.}
\label{fig:struct}
\end{figure}

The structures of $ {\rm CeRhSn} $ \cite{schmidt05} and $ {\rm CeRuSn} $
\cite{riecken07} are presented in Fig.\ \ref{fig:struct}
together with the coordination polyhedra of the cerium atoms. 
$ {\rm CeRhSn} $ (hexagonal ZrNiAl-type) has only one crystallographically 
independent cerium site with five rhodium, six tin, and six cerium atoms 
in the coordination shell. The cerium coordinations of the two 
crystallographically independent cerium sites in $ {\rm CeRuSn} $ 
(new monoclinic  type, space group $ C2/m $) have the same topology; 
however, small distortions lead to drastically different interatomic 
distances. Five ruthenium, seven tin, and four cerium atoms are within 
the coordination sphere of both cerium atoms. The striking structural 
features in both ternary stannides are short cerium-transition metal 
distances, which are 304-309\,pm Ce-Rh in $ {\rm CeRhSn} $, 
233-246\,pm Ce1-Ru and 288-291\,pm Ce2-Ru in $ {\rm CeRuSn} $, close 
to the sums of the covalent radii (290\,pm Ce + Rh and 289\,pm Ce + Ru),
\cite{emsley99} indicating strong Ce-Rh and Ce-Ru bonding. The shortest 
interatomic distances in both structures occur for Rh-Sn 
(277-285\,pm) and Ru-Sn (265-290\,pm), which compare well with the 
sums of the covalent radii (265\,pm Rh + Sn and 264\,pm Ru + Sn). Thus, 
the Ce-T and T-Sn interactions play the dominant role in chemical 
bonding in $ {\rm CeRhSn} $ and $ {\rm CeRuSn} $. In the present work, 
we perform a comparative study of the chemical bonding in these two 
compounds as based on density functional {\em ab initio} 
calculations. In doing so, we take into account previous calculations 
as well as experimental XPS studies on $ {\rm CeRhSn} $. 
\cite{schmidt05,slebarski02,slebarski04,shimada06,stewart01} 
Our calculations are based on the single-crystal data given in Ref.\ 
\cite{schmidt05,riecken07}. 

\section{Theoretical framework}
As a matter of fact, the degree of delocalization of 
the cerium $ 4f $ states depends on the applied pressure as well as on 
the environment in the crystal. In electronic structure calculations this 
delicate situation is addressed through various approaches treating 
the $ 4f $ states either as atomic like core states or as part of the 
valence basis set. This duality was experimentally 
evidenced in a combined analysis of $\mu$SR (muon spin relaxation) and 
neutron experiments on cerium intermetallic systems, which reveals the 
existence of magnetic excitations due to both conduction electrons at 
the Fermi level and well localized $ f $-electrons. \cite{yaouanc99}
Regarding the local environment, the quantum mixing (hybridization) 
of the $ 4f $ states with those of the ligand states can have large 
effects as well. This involves chemical bonding properties, which 
depend on the crystal structure as it is illustrated by the compounds 
under study.
 
\subsection{Computational method}
For the electronic structure calculations we used the augmented spherical 
wave (ASW) method in its scalar-relativistic implementation 
\cite{wkg,revasw,bookasw}. In the ASW method the wave function is 
expanded in atom-centered augmented spherical waves, which are Hankel 
functions and numerical solutions of Schr\"odinger's equation, 
respectively, outside and inside the so-called augmentation spheres. 
In order to optimize the basis set, additional augmented spherical waves 
were placed at carefully selected interstitial sites. The choice of these 
sites as well as the augmentation radii were automatically determined 
using the sphere-geometry optimization algorithm. \cite{sgo}
All valence states, including the Ce $ 4f $ orbitals, were treated as 
band states. In the minimal ASW basis set, we chose the outermost shells 
to represent the valence states using partial waves up to $ l = 4 $ for 
Ce as well as $ l= 3 $ for Ru, Rh and Sn. The completeness of the 
valence basis set was checked for charge convergence. The self-consistent 
field calculations are run to a convergence of $ \Delta Q=10^{-8} $\,Ryd 
\cite{mixpap} and the accuracy of the method is in the range of about 
1\,meV regarding energy differences.

\subsection{Spin-dependent calculations}
Due to the intermediate valent and trivalent natures identified for 
cerium within $ {\rm CeRuSn} $ we carried out spin-degenerate 
non-magnetic as well as spin-polarized calculations. Assuming 
a non-magnetic configuration means that the spin degeneracy is 
enforced for all species. 
Of course, such a configuration should not be confused with a paramagnet, 
which could be simulated either by a supercell calculation with random 
spin orientations or by calling for disordered local moment approaches  
based on the coherent potential CPA approximation \cite{niklasson03} or 
the LDA+DMFT scheme. \cite{held03,held06}

Subsequent spin-polarized calculations with different initial spin 
populations can lead at self-consistency either to finite or zero local 
moments within an implicit long range ferromagnetic order. They allow 
to confirm trends established within the mean-field analysis. 
Antiferromagnetic (AF) calculations can be carried out to test for the 
AF ground state. This can be done for instance  by enforcing an up-spin 
orientation for half of the crystallographic sites and a down-spin 
orientation for the other half thus creating two magnetic substructures. 
Such a procedure is followed here to identify the AF ground state of 
$ {\rm CeRuSn} $, which contains four formula units per cell.

\subsection{Assessment of chemical bonding properties}

To extract more information about the nature of the interactions 
between the atomic constituents from electronic  structure calculations, 
the crystal orbital overlap population (COOP) \cite{hoffmann87} or the 
crystal orbital Hamiltonian population (COHP) \cite{dronskowski93} may 
be employed. The latter has been already used to study chemical bonding 
in $ {\rm CeRhSn} $. \cite{schmidt05} While both the COOP and COHP 
approaches provide a qualitative description of the bonding, nonbonding, 
and antibonding interactions between two atoms, the COOP description 
in some cases exaggerates the magnitude of antibonding states.  A slight 
refinement was recently proposed in form of the so-called covalent bond 
energy $ E_{cov} $, which combines the COHP and COOP to calculate 
quantities independent of the particular choice of the potential zero. 
\cite{bester01} In the present work this covalent bond energy was used 
for the chemical bonding analysis. In the plots, negative, positive 
and zero magnitudes of $ E_{cov} $ are indicative of bonding, antibonding, 
and nonbonding interactions respectively.

\section{Results for non magnetic configurations}

\subsection{Site-projected density of states}

\begin{figure}[htbp]
\begin{center}
\subfigure{\includegraphics[width=\columnwidth]{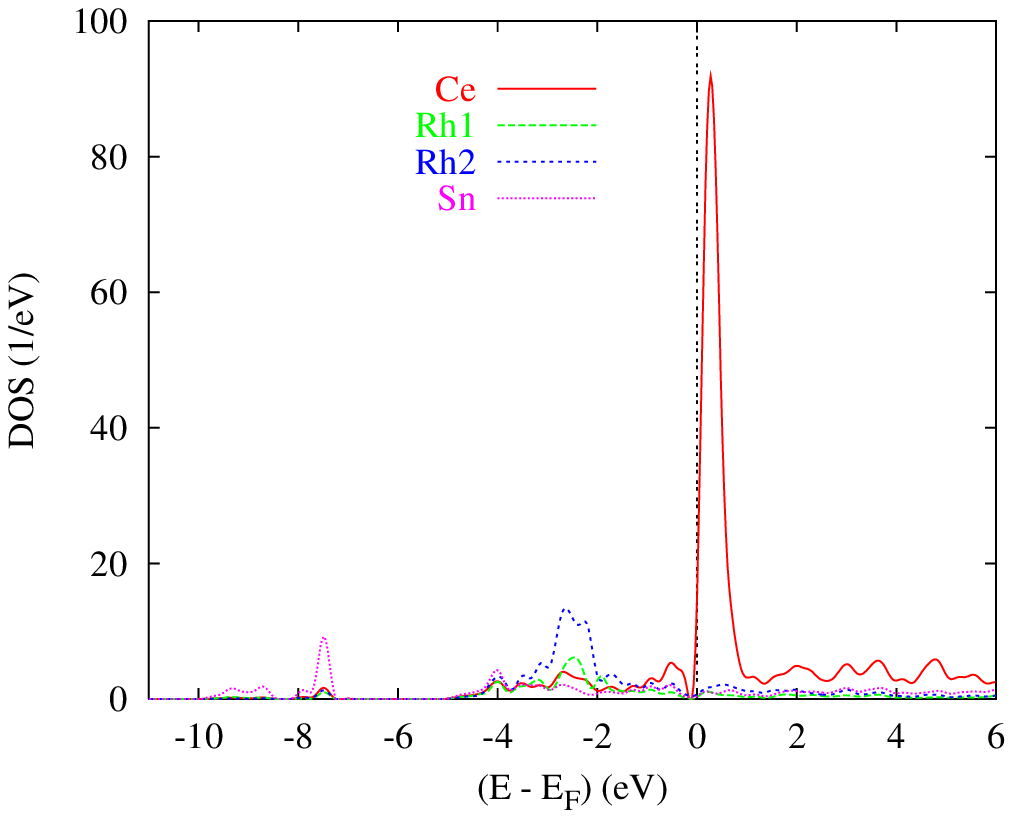}}
\subfigure{\includegraphics[width=\columnwidth]{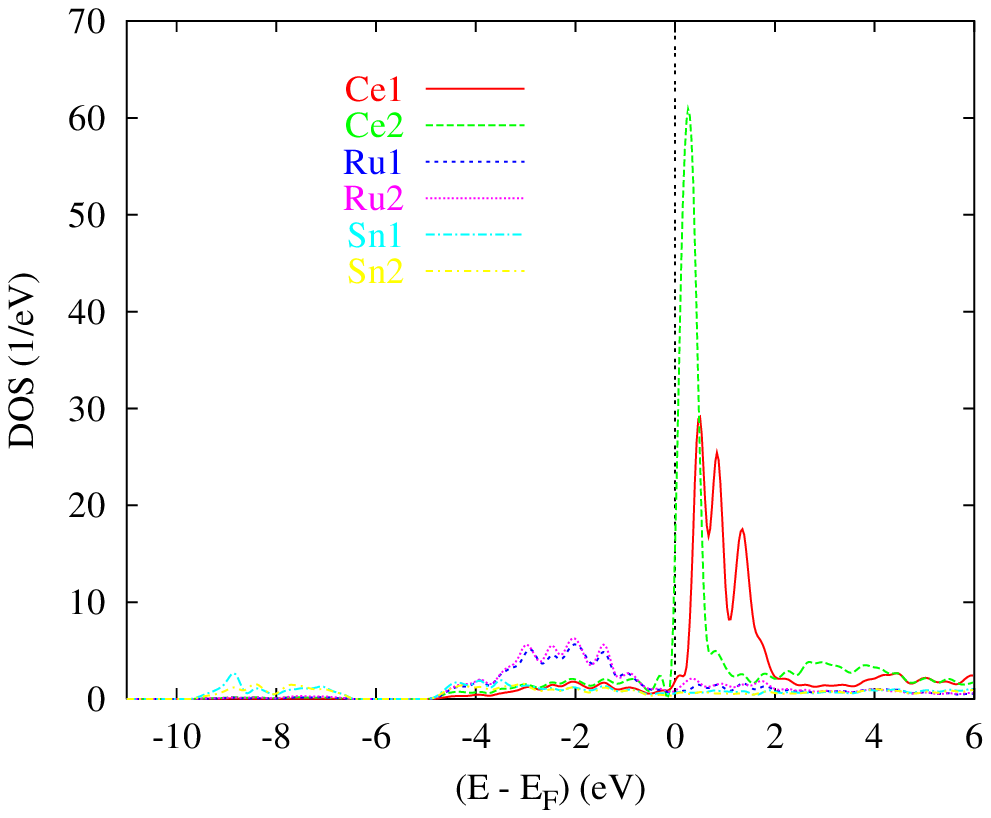}}
\caption{Non magnetic site projected DOS of $ {\rm CeRhSn} $ (a)  and 
         $ {\rm CeRuSn} $ (b)}
\label{fig1}
\end{center}
\end{figure}

The site-projected DOS (PDOS) for the two ternary stannides 
$ {\rm CeRhSn} $ and $ {\rm CeRuSn} $ are given in Fig. \ref{fig1}. 
In all panels the Fermi level ($ E_F $) is taken as zero energy. The 
cerium DOS are seen to prevail through the large peak around $ E_F $ 
mainly due to the $ 4f $ states. The PDOS of $ {\rm CeRhSn} $ show 
similarities with the plots obtained from the previous calclations 
\cite{schmidt05,slebarski02}, mainly for the itinerant part in the 
energy range from -4\,eV to $ E_F $, for which Ce XPS spectra were 
measured. \cite{slebarski02} However, major differences occur for 
$ {\rm CeRuSn} $, which contains two cerium sites. Broadened Ce1 PDOS 
are found mainly above the Fermi level with a small contribution at 
$ E_F $. In contrast, the Ce2 PDOS are localized at $ E_F $, which 
crosses the lower energy part of the Ce $ 4f $ states, and behave 
similarly to the Ce $ 4f $ states in $ {\rm CeRhSn} $. As compared 
to the latter, the larger splitting of the PDOS, mainly Ce1 as well 
as Ce2 is due to the lower symmetry of the monoclinic structure of the 
Ru-based system. There is a non negligible contribution from Ce 
itinerant states below $ E_F $ which ensure for the chemical bonding 
through the hybridization with the transition metal (Ru, Rh) $ 4d $ 
and Sn $ 5s $ and $ 5p $ states. Due to the large filling of their 
$ d $-states, the Ru and Rh PDOS are found below $ E_F $, completely 
within the valence band (VB). The different PDOS show similar shapes 
within the VB, mainly in the range from -5\,eV up to $ E_F $, which is 
indicative of the hybridization of the valence states involving 
itinerant Ce states below $ E_F $. We will return to this issue while 
discussing the chemical bonding in terms of the covalent bond energy 
$ E_{cov} $ below. 

\subsection{Analysis of the DOS within Stoner theory}

In as far as the Ce $ 4f $ states were treated as band states in the 
framework of our calculations the Stoner theory of band ferromagnetism
\cite{bookmat} can be applied to address the spin polarization at the 
different cerium sites. At zero temperature (ground state) one can 
express the total energy of the spin system resulting from the 
exchange and kinetic energies counted from a non-magnetic state as 
$ E= \frac{1}{2} [\frac{m^2}{n(E_F)}][1- I n(E_F)] $. 
Here $ I $ is the Stoner exchange-correlation integral, which is an 
atomic quantity derived from spin-polarized scalar-relativistic 
calculations \cite{brooks93}, and $ n (E_{F}) $ is the DOS at 
$ E_F $ in the non-magnetic state. From this expression, the product 
$ I n (E_F) $ provides a criterion for the stability of the spin system: 
A spin-polarized configuration (unequal spin occupation) will be more 
favourable if $ I n (E_F) > 1 $. The system then stabilizes through 
a gain of exchange energy. From the calculations we have for 
$ n (E_F) $ for Ce1, Ce2 in $ {\rm CeRuSn} $ and Ce in the rhodium 
based stannide 11, 98, and 39 states per Ryd, respectively. To analyze 
these results we use the Stoner integral for Ce as obtained from former 
scalar-relativistic calculations \cite{matar00}: 
$ I ({\rm Ce} 4f) \sim 0.02 $\,Ryd. The resulting Stoner products 
$ I n (E_F) $ for Ce1, Ce2 and Ce are, respectively, 0.22, 1.96 and 
0.78. The Stoner criterion is obeyed only for Ce2, which should carry 
a magnetic moment within $ {\rm CeRuSn} $ when spin polarized calculations 
are carried out. This agrees with its trivalent character versus the 
intermediate valent character of Ce1. Still, the rather large magnitude 
of $ I n (E_F) $ for Ce in $ {\rm CeRhSn} $ points to a tendency towards 
a magnetic instability. However, according to the experiment $ {\rm CeRhSn} $ 
is an intermediate-valence ternary stannide. For this reason, no ordered 
magnetic moment should develop when spin polarization is allowed for. 
This is further checked with the spin-polarized calculations outlined 
in the following section.

\subsection{Covalent bond energy $ E_{cov} $}

Chemical bonding properties can be already addressed on the basis of 
the spin-degenerate calculations. This is due to the fact that the 
spin-polarized bands, to a large degree, result from the spin-degenerate 
bands by a rigid energy shift. Negative, positive and zero values of 
$ E_{cov} $ indicate bonding, antibonding and nonbonding states,  
respectively. 

\subsubsection{$ {\rm CeRhSn} $} 

\begin{figure}[htbp]
\begin{center}
\includegraphics[width=\columnwidth]{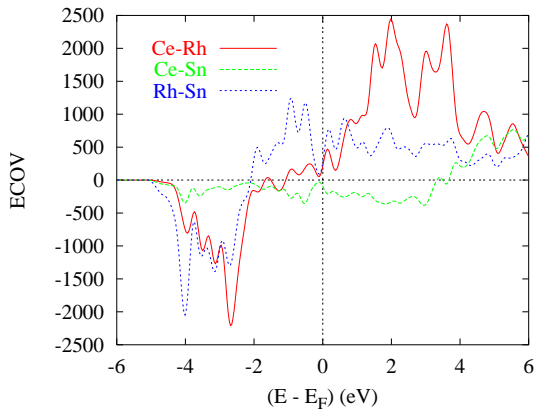}
\caption{Chemical bonding: non magnetic $ E_{cov} $ for $ {\rm CeRhSn} $.}
\label{fig2}
\end{center}
\end{figure}

The covalent bond energies $ E_{cov} $ for the different pairs of orbitals 
within $ {\rm CeRhSn} $ are shown in Fig.\ \ref{fig2}.  
Here, we have combined contributions of both Rh sites in order to keep 
the representation simple. Again, energies are referred to the Fermi 
level. Note that the $ E_{cov} $ intensities along the ordinate are 
unitless and should be considered at a qualitative level. From the 
bottom of the valence band (VB) up to the  Fermi level the dominant 
interactions result from the Ce-Rh and Rh-Sn bonds, which are found strongly 
bonding up to -2\,eV. While the Ce-Rh bond has a rather nonbonding 
character up to $ E_F $, it only starts to be antibonding above $ E_F $ 
within the empty conduction band (CB). One can also notice an antibonding 
peak above $ E_F $, which can also be observed for Ce2-Ru2 in 
$ {\rm CeRuSn} $ (see Fig.\ \ref{fig3}) as it will be shown below. To the 
contrary, antibonding Rh-Sn are observed early within the VB, which 
destablizes the system. To conclude, the Ce-Rh bond  is evidently the 
stabilizing contribution, followed by the Ce-Sn one, which is of 
smaller magnitude. It is interesting to note that, although the definition 
of the covalent bond energy $ E_{cov} $ is somewhat different from that 
of the COHP, the same bonding trends of $ {\rm CeRhSn} $ are obtained 
from the calculations of Schmidt {\em et al.} using the latter. 
\cite{schmidt05}

\subsubsection{$ {\rm CeRuSn} $} 

\begin{figure}[htbp]
\begin{center}
\subfigure{\includegraphics[width=\columnwidth]{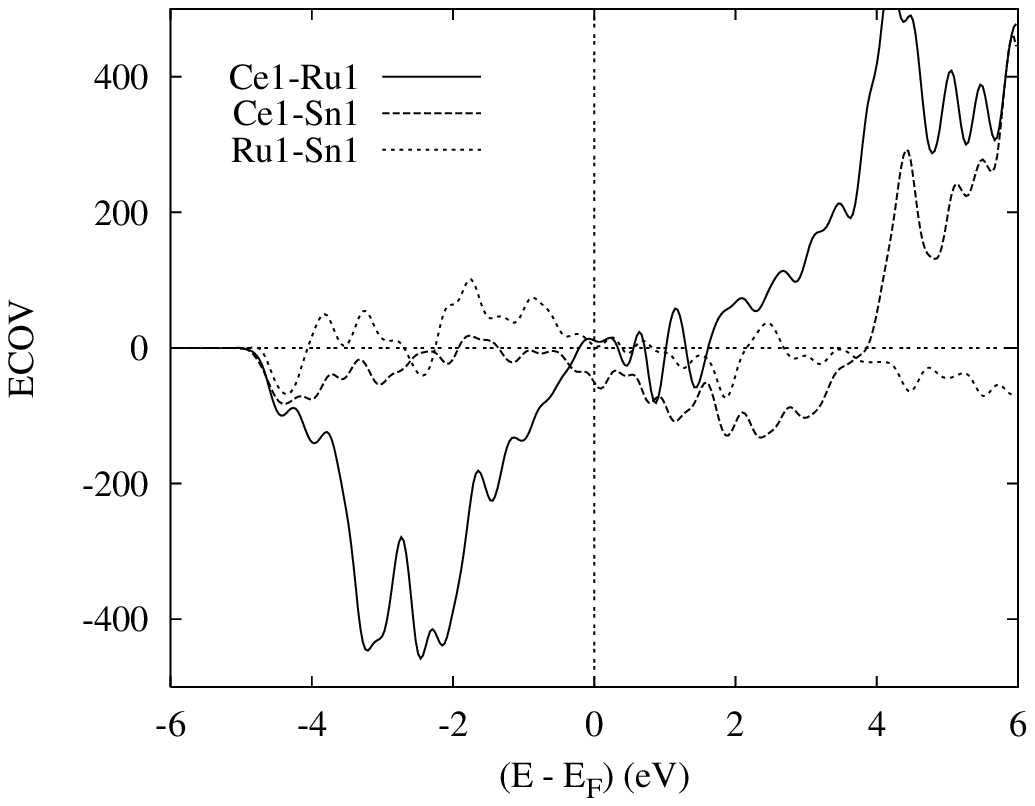}}
\subfigure{\includegraphics[width=\columnwidth]{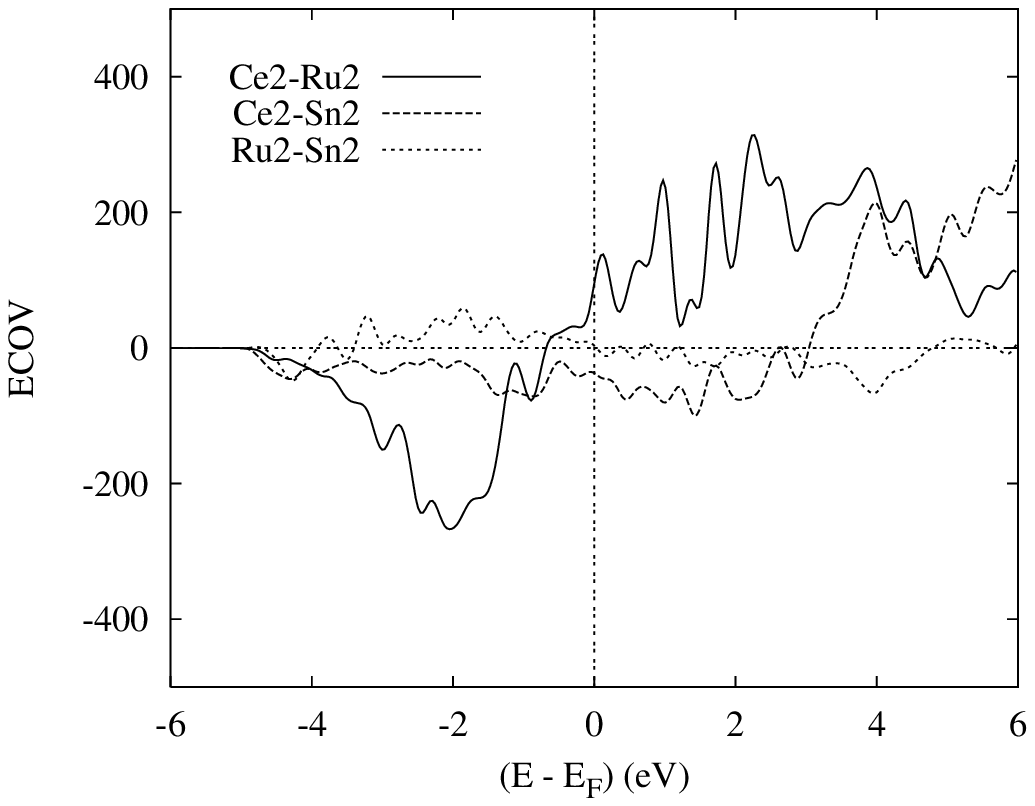}}
\caption{Chemical bonding: non magnetic $ E_{cov} $ for $ {\rm CeRuSn} $ 
         in two subcells.}
\label{fig3}
\end{center}
\end{figure}

Due to the superstructure of this system it is useful to study the 
interactions within two subsystems separately. The result is presented 
in Fig.\ \ref{fig3}. 
There, only one atom of each kind is involved. As a consequence, the 
absolute $ E_{cov} $ values are smaller than in Fig.\ \ref{fig2}. The 
major interactions in the two subsystems occur clearly for Ce-Ru, which 
is bonding throughout the VB but with twice larger intensity for Ce1 
than for Ce2. This follows from the relative distances whereby 
$ 233 < d_{Ce1-Ru} < 246 $\,pm while 
$ 288 < d_{Ce2-Ru} < 291 $\,pm. This is concomitant with the different 
behavior of the two cerium sites as discussed above. The larger Ce2-Ru 
separation leads to a stronger localization of Ce2 states, which fact 
is favorable for the onset of an atomic magnetic moment. We also note 
that the Ce-Rh separation is even larger ($ d_{Ce-Rh} \sim 307 $\,pm) in 
$ {\rm CeRhSn} $. The antibonding peak above the Fermi level just 
like in the Rh based system is in line with the magnetic instability 
of Ce2.  Finally, albeit with a much smaller magnitude, the Ce-Sn 
interaction is bonding throughout the VB and should contribute to the 
bonding within the system.

\section{Results of the spin polarized configurations}

Spin-polarized calculations for the magnetic structures were carried 
out by initially allowing for two different spin occupations, then 
self-consistently converging the charges and the magnetic moments. 
We assume firstly a hypothetic ferromagnetic configuration without 
any constraint on the spins. Then antiferromagnetic computations were 
carried out for $ {\rm CeRuSn} $ in order to identify the ground state 
from energy differences.

\subsection {$ {\rm CeRhSn} $}  

Independent of the experimental finding of the intermediate valent 
character of Ce within $ {\rm CeRhSn} $, it was suggested from our 
spin-degenerate calculations as well as from Ref.\ \cite{slebarski02} 
that the system could be on the verge of a magnetic instability. However, 
we could not identify a finite magnetic moment on Ce from self-consistent 
computations at high precision sampling of $ {\bf k} $-space in full 
agreement with the experiments.

\subsection {$ {\rm CeRuSn} $}  

\begin{figure}[htbp]
\begin{center}
\includegraphics[width=\columnwidth]{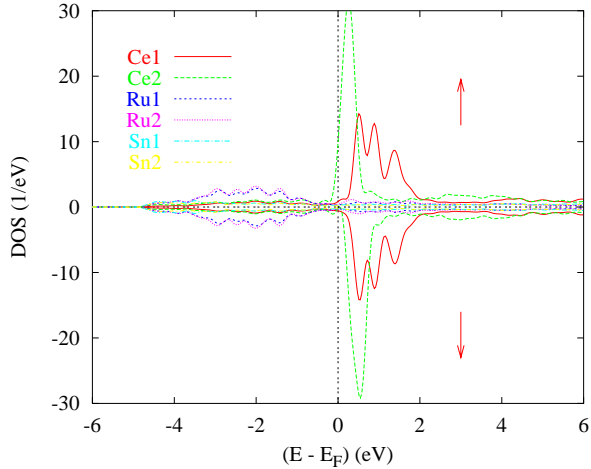}
\caption{Site and spin projected DOS of $ {\rm CeRuSn} $ in the 
         ferromagnetic hypothetic state.}
\label{fig4}
\end{center}
\end{figure}

Contrary to the former system, a stable magnetic solution was identified 
at high precision BZ sampling. The energy difference favoring the 
ferromagnetic state is $ \Delta E =7.14 $\,meV per formula unit.  
The spin-only moments were $ M({\rm Ce1})=0.005 \mu_{B} $, 
$ M({\rm Ce2})= 0.44 \mu_{B} $, $ M({\rm Ru1})=-0.015 \mu_{B} $ and 
$ M({\rm Ru2})=-0.041 \mu_{B} $. While the vanishingly small moment of Ce1 
resembles the results of $ {\rm CeRhSn} $, the finite moment on Ce2 
is in agreement with the experimental finding as to the trivalent 
behavior of cerium. The magnetic moments on Ru1 and Ru2 are of induced 
nature through the hybridization of their states with the respective 
Ce1 and Ce2 states. These results are illustrated at Fig.\ \ref{fig4} 
showing the site and spin projected density of states of 
$ {\rm CeRuSn} $. Low lying Sn $ 5s $ states are not shown and the 
energy window from $ -6 $ to 6\,eV shows the different behavior of 
the different Ce sites, whereby the exchange splitting is observed 
only for Ce2. The main bonding characteristics follow  the disucssion 
above of the non magnetic DOS.
 
Spin-orbit coupling effects can be large in Ce based magnetic systems. 
This is because the localized character of the $ 4f $ wave function 
leads to the formation of orbital moments. We use the orbital field 
(OR) scheme introduced by Brooks \cite{brooks93} as well as Sandratskii 
and K\"{u}bler \cite{sandratskii95}, which helped to account for the 
experimental moments within formerly studied Ce intermetallic systems 
\cite{matar00,matar07}. The trend is that the magnitude of the orbital 
moment (L) is close to that of the spin-only (SO) moment but with 
opposite signs, in agreement with Hund's 3rd rule. The L moment 
of cerium stems from a $ 4f $(Ce2) occupation of $\sim 1.3 $ electrons  
whose orbital moment ($ \sim 2.5 \mu_{B} $) comes close to the one of 
an atomic orbital, namely $ 3 \mu_{B} $ as expected from Hund's 
2nd rule. This reflects an atomic-like character of the $ 4f $ (Ce) 
subshell. A resulting ordered LS moment of cerium of $ \sim 2 \mu_B $ 
is then obtained; it should be confronted with experimental magnitudes 
of magnetization from neutron diffraction when the results are made 
available. 

Lastly, in order to check for the ground state, AF calculations were 
carried out. At self consistency, $ \Delta E = -2.3 $\,meV in favor of 
antiferromagnetic ordering was obtained thus pointing to an 
antiferromagnetic ground state in agreement with the experiment. The 
moment carried by Ce2 is $ \pm 0.39 \mu_B $, slightly smaller than 
in the ferromagnetic configuration. This is likely due to the decrease 
of symmetry when the two magnetic substructures were accounted for within 
the unit cell.

\section{Conclusion}

In this work we have undertaken theoretical investigations of the two 
ternary equiatomic cerium stannides $ {\rm CeRhSn} $ and 
$ {\rm CeRuSn} $ as based on accurate X-ray determinations. 
Calculated densities of states and chemical bonding properties for 
$ {\rm CeRhSn} $ agree very well with former theoretical results. 
$ {\rm CeRuSn} $ is strongly influenced by the peculiar feature of 
a complex interplay of both intermediate valent and trivalent Ce atoms. 
The exciting properties of this compound were addressed both by 
spin-degenerate and spin-polarized density functional theory based 
calculations. Analysis of the electronic structures and of the chemical 
bonding reveal different types of chemical bonds due to the nature of 
the Ce site and its environment with Ru, Sn as well as Ce ligands. As a 
consequence, a local magnetic moment is expected only for the 
cerium site with trivalent character. Indeed, spin-polarized 
calculations lead to a nearly vanishing magnetic moment at the 
intermediate-valent Ce1 site and a finite moment shows up only at 
the trivalent Ce2 site. In agreement with experimental data, the 
Ce2 moments give rise to an antiferromagnetic ground state.

\begin{acknowledgments}
Fruitful discussions with E.-W.\ Scheidt are gratefully acknowledged. 
Computational facilities were provided within the intensive numerical 
simulation facilities network M3PEC of the University Bordeaux 1 partly 
financed by the ``Conseil R\'egional d'Aquitaine''. 
This work was supported by the Deutsche Forschungsgemeinschaft through
SFB 484 and SCHE /7-1.
\end{acknowledgments}


\begin{thebibliography}{}
\bibitem{takabatake87} 
T.\ Takabatake, Y.\ Nakazawa, and M.\ Ishikawa, 
Jpn.\ J.\ Appl.\ Phys.\ Suppl.\ {\bf 26}, 547 (1987).

\bibitem{schmidt05} 
T.\ Schmidt, D.\ Johrendt, C.\ P.\ Sebastian, R.\ P\"ottgen, K.\ L\c{a}tka, 
and R.\ Kmie\'c, 
Z.\ Naturforsch.\ {\bf 60b}, 1036 (2005).

\bibitem{chevalier06} 
B.\ Chevalier, C.\ P.\ Sebastian, and R.\ P\"ottgen, 
Solid State Sci.\ {\bf 8}, 1000 (2006).

\bibitem{adroja88} 
D.\ T.\ Adroja, S.\ K.\ Malik, B.\ D.\ Padalia, and R.\ Vijayaraghavan, 
Solid State Commun.\ {\bf 66}, 1201 (1988).

\bibitem{riecken05} 
J.\ F.\ Riecken, G.\ Heymann, T.\ Soltner, R.-D.\ Hoffmann, H.\ Huppertz, 
D.\ Johrendt, and R.\ P\"ottgen, 
Z.\ Naturforsch.\ {\bf 60b}, 821 (2005).

\bibitem{baran97} 
S.\ Baran, J.\ Leciejewicz, N.\ St\"usser, A.\ Szytu{\l}a, A.\ Zygmunt, 
and Y.\ Ding, 
J.\ Magn.\ Magn.\ Mater.\ {\bf 170}, 143 (1997).

\bibitem{lenkewitz96} 
M.\ Lenkewitz, S.\ Cors\'epius, and G.\ R.\ Stewart, 
J.\ Alloys Comp.\ {\bf 241}, 121 (1996).

\bibitem{scifinder} 
216, 45, and 51 entries occur for the formulae $ {\rm CeNiSn} $, 
$ {\rm CePdSn} $, and $ {\rm CePtSn} $ in the SciFinder Scholar, 
version 2007 (http://www.cas.org/SCIFINDER/SCHOLAR/)

\bibitem{riecken07} 
J.\ F.\ Riecken, W.\ Hermes, B.\ Chevalier, R.-D.\ Hoffmann, 
F.\ M.\ Schappacher, and R.\ P\"ottgen, 
Z.\ Anorg.\ Allg.\ Chem.\ {\bf 633}, 1094 (2007).

\bibitem{hoffmann07} 
R.-D.\ Hoffmann, B.\ Chevalier, J.\ F.\ Riecken, U.\ Ch.\ Rodewald, 
F.\ M.\ Schappacher, and R.\ P\"ottgen, 
unpublished results.

\bibitem{hohenberg64}
P.\ Hohenberg and W.\ Kohn, 
Phys.\ Rev.\ B, {\bf 136}, 864 (1964). 

\bibitem{kohn65} 
W.\ Kohn and L.\ J.\ Sham, 
Phys.\ Rev.\ A, {\bf 140}, 1133 (1965). 

\bibitem{wkg} 
A.\ R.\ Williams, J.\ K\"{u}bler and C.\ D.\ Gelatt,
Phys.\ Rev.\ B {\bf 19}, 6094 (1979).\ 

\bibitem{revasw}
V.\ Eyert, 
Int.\ J.\ Quantum Chem.\ {\bf 77}, 1007 (2000). 

\bibitem{bookasw}
V.\ Eyert, 
{\em The Augmented Spherical Wave Method -- A Comprehensive Treatment}, 
Lect.\ Notes Phys.\ {\bf 719} (Springer, Berlin Heidelberg 2007).

\bibitem{emsley99} 
J.\ Emsley, 
{\em The Elements} (Oxford University Press, Oxford, 1999).

\bibitem{slebarski02} 
A.\ Slebarski, M.\ Radlowska, T.\ Zawada, M.\ B.\ Maple, 
A.\ Jezierski, and A.\ Zygmunt, 
Phys.\ Rev.\ B {\bf 66}, 104434 (2002).

\bibitem{slebarski04} 
A.\ Slebarski, T.\ Zawada, J.\ Spalek, and A.\ Jezierski, 
Phys.\ Rev.\ B {\bf 70}, 235112 (2004).

\bibitem{shimada06} 
K.\ Shimada, H.\ Namatame, M.\ Taniguchi, M.\ Higashiguchi, S.-I.\ Fujimori, 
Y.\ Saitoh, A.\ Fujimori, M.\ S.\ Kim, D.\ Hirata, and T.\ Takabatake, 
Physica B {\bf 791}, 378 (2006). 

\bibitem{stewart01}
G.\ R.\ Stewart, 
Rev.\ Mod.\ Phys.\ {\bf 73}, 797 (2001).

\bibitem{yaouanc99} 
A.\ Yaouanc, P.\ Dalmas de Réotier, P.\ C.\ M.\ Gubbens, C.\ T.\ Kaiser, 
P.\ Bonville, J.\ A.\ Hodges, A.\ Amato, A.\ Schenck, P.\ Lejay, 
A.\ A.\ Menovsky and M.\ Mihalik.\  
Physica B {\bf 259-261}, 126 (1999).

\bibitem{sgo} 
V.\ Eyert and K.-H.\ H\"ock, 
Phys.\ Rev.\ B {\bf 57}, 12727 (1998).

\bibitem{mixpap}
V.\ Eyert, 
J.\ Comput.\ Phys.\ {\bf 124}, 271 (1996).

\bibitem{niklasson03} 
A.M.N.\ Niklasson, J.M.\ Wills, M.I.\  Katsnelson, I.A.\ Abrikosov, 
O.\ Eriksson, and B.\ Johansson, 
Phys.\ Rev.\ B {\bf 67}, 235105 (2003).

\bibitem{held03}
K.\ Held, V.\ I. Anisimov, V.\ Eyert, G.\ Keller, A.\ K.\ McMahan,
I.\ A.\ Nekrasov, and D.\ Vollhardt, 
Adv.\ Solid State Phys.\ {\bf 43}, 267 (2003).

\bibitem{held06}
K.\ Held, I.\ A.\ Nekrasov, G.\ Keller, V.\ Eyert, N.\ Bl\"umer,
A.\ K.\ McMahan, R.\ T.\ Scalettar, T.\ Pruschke, V.\ I.\ Anisimov,
and D.\ Vollhardt, 
phys.\ stat.\ sol.\ (b) {\bf 243}, 2599 (2006).

\bibitem{hoffmann87} 
R.\ Hoffmann, 
Angew.\ Chem.\ Int.\ Ed.\ Engl.\ {\bf 26}, 846 (1987). 

\bibitem{dronskowski93} 
R.\ Dronskowski and P.\ E.\ Bl\"{o}chl, 
J.\ Phys.\ Chem.\ {\bf 97}, 8617 (1993). 

\bibitem{bester01} 
G.\ Bester and M.\ F\"ahnle, 
J.\ Phys: Condens.\ Matter {\bf 13}, 11541 (2001). 

\bibitem{bookmat} 
J.\ K\"ubler and V.\ Eyert, 
{\em Electronic structure calculations}, 
in: {\em Electronic and Magnetic Properties of Metals and Ceramics},
    edited by K.\ H.\ J.\ Buschow
    (VCH Verlagsgesellschaft, Weinheim, 1992), pp.\ 1-145; 
    Volume 3A of
    {\em Materials Science and Technology},
    edited by R.\ W.\ Cahn, P.\ Haasen, and E.\ J.\ Kramer
    (VCH Verlagsgesellschaft, Weinheim, 1991-1996).

\bibitem{brooks93} 
M.\ S.\ S.\ Brooks and B.\ Johansson, 
in: {\em Handbook of Magnetic Materials}, ed.\ K.\ H.\ J.\ Buschow, 
    Vol.\ {\bf 7} (1993). 

\bibitem{sandratskii95} 
L.M.\ Sandratskii and J.K\"ubler, 
Phys.\ Rev.\ Lett.\ {\bf 75}, 946 (1995).

\bibitem{matar00} 
S.\ F.\ Matar and A.\ Mavromaras,
J.\ Solid State Chem.\ {\bf 149}, 449 (2000). 

\bibitem{matar07} 
S.\ F. Matar, 
Phys.\ Rev.\ B {\bf 75}, 104422 (2007).

\end{thebibliography}
\end{document}